%% file: sample-sigconf.tex
\documentclass[sigconf]{acmart}

\usepackage{booktabs} 
\usepackage{multirow}
\usepackage{scalefnt}
\usepackage{tikz}
\usepackage[size=small]{caption}

\setcopyright{rightsretained}

\makeatletter
\renewcommand\@formatdoi[1]{\ignorespaces}
\makeatother
 
\acmISBN{}

\acmConference[RecSys'18]{the Late-Breaking Results track part of the Twelfth ACM Conference on Recommender Systems}{October 2-7, 2018}{Vancouver, BC, Canada}

\acmYear{2018}
\copyrightyear{2018}
\acmPrice{15.00}
\begin{document}

\title{CoBaR: Confidence-Based Recommender}

\author{Fernando S. Aguiar Neto, Arthur F. da Costa and Marcelo G. Manzato}
\affiliation{%
\institution{Institute of Mathematical and Computer Sciences - University of S\~ao Paulo, S\~ao Carlos, SP, Brazil}
\institution{fernando.soares.aguiar@usp.br, fortes@icmc.usp.br and mmanzato@icmc.usp.br}
}

\begin{abstract}
    Neighborhood-based collaborative filtering algorithms usually adopt a fixed neighborhood size for every user or item, although groups of users or items may have different lengths depending on users' preferences. In this paper, we propose an extension to a non-personalized recommender based on confidence intervals and hierarchical clustering to generate groups of users with optimal sizes. The evaluation shows that the proposed technique outperformed the traditional recommender algorithms in four publicly available datasets.
\end{abstract}

\keywords{\small Recommender Systems; Confidence Interval;  Hierarchical Clustering}

\maketitle

\input{samplebody-conf}

\bibliographystyle{ACM-Reference-Format}
\bibliography{recsys}
\footnotesize

\end{document}

%% file: samplebody-conf.tex
\section{Introduction}

Recommending most popular items is a very simple and effective way to provide non-personalized recommendations to users. It consists of using the global mean rating for each item, and selecting those items with highest ratings to be recommended. One approach to personalize most popular is to reduce the scope of aggregated ratings, similarly to User-$k$NN and Item-$k$NN algorithms \cite{Aggarwal:2016}, where predictions are calculated using information from the $k$ most similar users or items, respectively. One drawback of these models is that the neighborhood size is fixed, which may reduce the algorithm's accuracy as the number of individuals with similar preferences considered by the system is not optimal. Some works ~\cite{Herlocker2002, Adamopoulos2014} studied the effects of neighborhood size in collaborative filtering, but usually the length of clusters is not personalized for each user. On the other hand, hierarchical clustering has been applied in recommender systems to achieve variable neighborhood sizes \cite{NAJAFABADI2017113}; however, choosing the best cluster for each user in the hierarchical structure is a challenging task that needs better research. 

Motivated by the simplicity of most popular and neighborhood-based algorithms aforementioned, we propose CoBaR, an approach to limit the range in which global mean of ratings is inferred. We use a hierarchical clustering technique to generate users' groups, and for each user-item pair, we find the set of users in which the ratings given to that specific item are similar, using the concept of confidence interval of the item's ratings in each group. In this way, our approach aims to improve non-personalized techniques, such as most popular, by setting optimal hubs of like-minded users.

This paper is structured as follows: in Section \ref{sec:ProposedApproach} the proposed approach is detailed; in Section \ref{sec:MethodologyAndEvaluation} the experimental methodology and results evaluation are presented and discussed; lastly, Section \ref{sec:FinalRemarks} addresses final remarks and future work.

\vspace{-8pt}

\section{Proposed Approach}
\label{sec:ProposedApproach}

In this work, we propose CoBaR, an approach to narrow the information used for non-personalized approaches, so that they achieve competitive results when compared to personalized algorithms, but with less complexity. For such, users are grouped in a hierarchy, using an agglomerative clustering solution~\cite{Gan2007DataApplications}. This approach generates users' groups of varied sizes, starting at a cluster with only one user, and aggregating new users until all belong to a single cluster. This variety in size is desirable, because it permits many neighborhoods in our approach. 

Given a cluster with an arbitrary number of users, it will contain only a subset of all interactions in the system, therefore the ratings provided to an item will vary from cluster to cluster. Our approach uses the concept of confidence intervals, which is well-known in statistics \cite{hinkley1979theoretical}. Given a sample, the mean and standard deviation are calculated, and an interval is inferred with certain confidence, e.g. 95\%. It means that if one takes another sample and generates another interval and repeats this process infinitely, 95\% of the intervals would contain the real mean of the studied population~\cite{hinkley1979theoretical}.

We assume that our samples lead to one of the 95\% confidence intervals in which the real mean belongs to. By assuming that, it is desirable that we find the set of samples which leads to the smallest confidence interval possible, providing little margin for error. For instance, let's say that we presume the real mean rating for an item $i$ is within $(1, 100)$. Even considering that this interval really contains the real mean, certainly it is a relatively long interval. If we manage to find an interval $(10,25)$ for the same item using other samples, the margin for error is smaller and therefore we assume the samples are better.

Notice that our intent is not to predict the real mean rating for an item on the entire dataset, but to predict which rating a specific active user would give to that item. Therefore, it makes sense to find the smallest interval because it indicates that for that item, the users similar to the active user tend to generate similar ratings.

Each generated group of users will provide a different sample of ratings for each item, thus a different confidence interval. In order to predict the rating $\hat{r}_{ui}$ that a user $u$ would give to an item $i$, we use the information of the group in which the confidence interval of the ratings given to item $i$ is the smallest and also contains user $u$. Once this cluster $c$ is found, $\hat{r}_{ui}$ is calculated as $\hat{r}_{ui} = \gamma b_u + (1 - \gamma)b^c_i$.

\noindent where $\gamma$ is a balance factor, $b_u$ is the mean rating given by the user in the entire dataset, and $b^c_i$ is the mean rating given to item $i$ in the optimal cluster $c$, i.e. the mean of the smallest confidence interval for the item $i$'s ratings in a group that contains user $u$. If an item has only one rating, no confidence interval can be inferred. In this case, we simply calculate $\hat{r}_{ui}$ as the mean rating given by the user.

\begin{figure}[!ht]
        \centering
        \includegraphics[draft=false, scale=0.5]{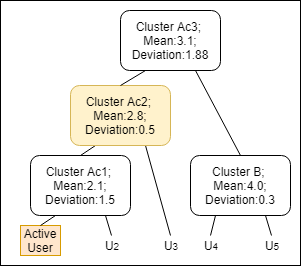}
        \caption{Example of hierarchy of users with the confidence intervals for item $i$. For each cluster, it is presented the cluster's label, the mean rating for item $i$ in that group and deviation, which indicates the distance that the bounds of the interval have to the mean. Clusters labelled Ac<x> are the ones containing the active user $u$.}
        \label{fig:GroupHierarchy}
\end{figure}

In order to illustrate the functioning of our approach, consider the users' clustering hierarchy presented in Figure \ref{fig:GroupHierarchy}. Besides the hierarchical structure, the figure also presents, for each group, the mean rating for an item $i$ and the size of the confidence interval. Notice that the active user joins $U_2$ in $Cluster Ac1$, and for item $i$ this group has a ratings deviation of $1.5$. On the other hand, when user $U_3$ joins the group in $Cluster Ac2$, the deviation turns to be $0.5$, which is the smallest for the groups that have the active user. Therefore, we have $b^c_i = 2.8$ and assuming the mean rating given by the active user to all items is $b_u = 3.4$, and $\gamma = 0.5$, then $\hat{r}_{ui} = (2.8 + 3.4)/2 = 3.1$.

\vspace{-8pt}

\vspace{-18pt}

\section{Methodology and Evaluation}
\label{sec:MethodologyAndEvaluation}

We base our evaluation on a comparison with the well-known recommender algorithms: User-$k$NN (UKNN), Item-$k$NN (IKNN) and Matrix Factorization (MF); and with Most Popular (MP), which is a non-personalized algorithm. We evaluate our approach in four public available datasets: CiaoDVD\footnote{https://www.librec.net/datasets.html}, FilmTrust$^1$, Booking Crossing\footnote{http://www2.informatik.uni-freiburg.de/~cziegler/BX/} and Amazon Digital Music\footnote{http://jmcauley.ucsd.edu/data/amazon/}. To measure the predictive accuracy of the different methods, we used the Root Mean Square Error measure (RMSE) \citep{Aggarwal:2016}, with $10$-fold cross-validation. We computed the mean across the $10$-folds and compared our method against the competitors using a Wilcoxon test with a $99\%$ confidence level~\citep{Fay:2010}. The baseline competitors belong to the Case Recommender Framework version 1.0.13\footnote{https://pypi.python.org/pypi/CaseRecommender}. We used the default framework settings for recommender algorithms and fixed $\gamma = 0.5$. 
The hierarchical clustering used was Ward's Method \cite{Gan2007DataApplications}, using Cosine distance among users, which were represented by their rating vectors.

Table \ref{tab:results} shows the results of this evaluation, for all datasets. We note that the proposed method, CoBaR, achieved statistically better results than the baselines, as proven by the Wilcoxon test analysis ($p$-value < 0.01). This indicates that the use of confidence intervals narrows users' ratings in a way better than traditional neighborhood and collaborative approaches. 

\begin{table}[!ht]
  \caption{Comparison between our proposed approach with baselines in terms of RMSE. Bold typeset indicates the best performance.}
  \vspace{-10pt}
  \label{tab:results}
  \begin{tabular}{cccccc}
    \toprule
    Datasets & CoBaR & UKNN & IKNN & MP & MF\\
    \midrule
    \texttt{CiaoDVD} & \textbf{0.955*} & 1.152 & 1.323 & 1.023 & 1.012\\
    \texttt{FilmTrust} & \textbf{0.823*} & 0.848 & 1.025 & 0.926 & 0.906\\
    \texttt{Booking Crossing}& \textbf{0.785*} & 0.867 & 0.970 & 0.878 & 0.832\\
    \texttt{Amazon D. Music}& \textbf{0.891*} & 0.941 & 1.049 & 0.981 & 0.936 \\
    \bottomrule
  \end{tabular}
  
  \vspace{-12pt}
\end{table}

According to the experiments, neighborhoods with varied sizes were useful to select the best ratings to be aggregated for each recommendation. Such results are competitive to related baselines, such as User and Item-$k$NN, Most Popular and Matrix Factorization, showing that more studies in hierarchical clustering and confidence intervals are promising for further research.

\vspace{-5pt}

\section{Final Remarks}
\label{sec:FinalRemarks}

This paper presented CoBaR, an approach to limit the scope of ratings based on smallest confidence intervals for each item ratings, providing a reasonable and intuitive approach for a rating prediction scenario. Using a hierarchical clustering pre-processing step, users can be better grouped according to their preferences, and a simple non-personalized algorithm can be applied for each user, producing competitive accuracy when compared to personalized techniques.  

We conducted experiments in four datasets from different domains (movies, books and music) and the results show that our strategy improves the overall system's accuracy. The main advantage of our approach is to provide a more guided recommendation for each user, which helps to mitigate the search space and computational cost problems in recommender systems. In addition, we are able to customize and optimize Most Popular Items, a non-personalized recommendation algorithm, with the personalized sets generated by our method.

In future work, we intend to apply this technique to different types of metadata and similarity metrics. Furthermore, we intend to use other approaches to navigate through the hierarchical structure, and use machine learning algorithms to find better values to $\gamma$ in order to improve the results.

\vspace{-5pt}

\begin{acks}
We would like to acknowledge FAPESP (numbers 2016/20280-6 and 2016/04798-5) and CAPES.

\end{acks}
\vspace{-5pt}